# The Butterfly Effect: Correlations Between Modeling in Nuclear-Particle Physics and Socioeconomic Factors

Maria Grazia Pia, Tullio Basaglia, Zane W. Bell and Paul V. Dressendorfer

***Abstract*–A scientometric analysis has been performed on selected physics journals to estimate the presence of simulation and modeling in physics literature in the past fifty years. Correlations between the observed trends and several social and economical factors have been evaluated.**

## I. Introduction

The development of models is exploited in physics research as a method to understand the origin of experimentally observed effects, or to predict them. Nowadays models describing particle interactions with matter and the resulting detector response are usually embedded in simulation tools, often based on Monte Carlo methods.

A scientometric analysis of particle and nuclear physics literature [1], presented at the 2009 IEEE Nuclear Science Symposium, showed the surprising result that more than 60% of the papers published in IEEE Transactions on Nuclear Science (TNS) in recent years mention modeling, and a large fraction of them mention simulation or Monte Carlo. Similar results were observed in journals with similar scope, such as Nuclear Instruments and Methods (NIM).

These preliminary observations have been investigated in greater depth. This paper reports detailed statistics regarding the trends of modeling and simulation in particle/nuclear physics literature, with emphasis on nuclear technology journals; the analysis spans a more extended time range than the period analyzed in [1]. It also investigates the correlation of these trends with economic factors, advances in computing technology, academic factors, and major evolutions in the nuclear and particle physics domain over the past decades.

The detailed analysis has been focused on TNS publications.

## II. Modeling and Simulation in Nuclear Physics Literature

Modeling and simulation play an important role in experimental research in nuclear and particle physics, and related fields. The contribution of these activities to the production of experimental results has been evaluated through a scientometric study of representative literature. The analysis estimated the fraction of articles which mention them, out of the total number of papers published each year, over a period of five decades (1960-2009). The data have been extracted from the journal publishers' web sites by means of the full-text search tools they provide.

An example is shown Fig. 1, which illustrates the trend of various full-text search patterns in TNS. A pattern of steadily increasing contribution of modeling and simulation to the production of results is clearly visible over the five decades analyzed in this study: the fraction of papers published in TNS that mention these words (or variants of them) has increased from a few percent in the early 60's to more than 50% in recent years.

Fig. 1 shows that the increased relative presence of modeling in TNS publications is mostly due to an increased popularity of simulation (including Monte Carlo simulation) over the years while the fraction of papers mentioning only modeling (or variants of this string) has been approximately stable, or slightly decreasing, over the past five decades. Therefore, the following analysis has been limited to articles mentioning "simulation" or "Monte Carlo" in the body of the text.

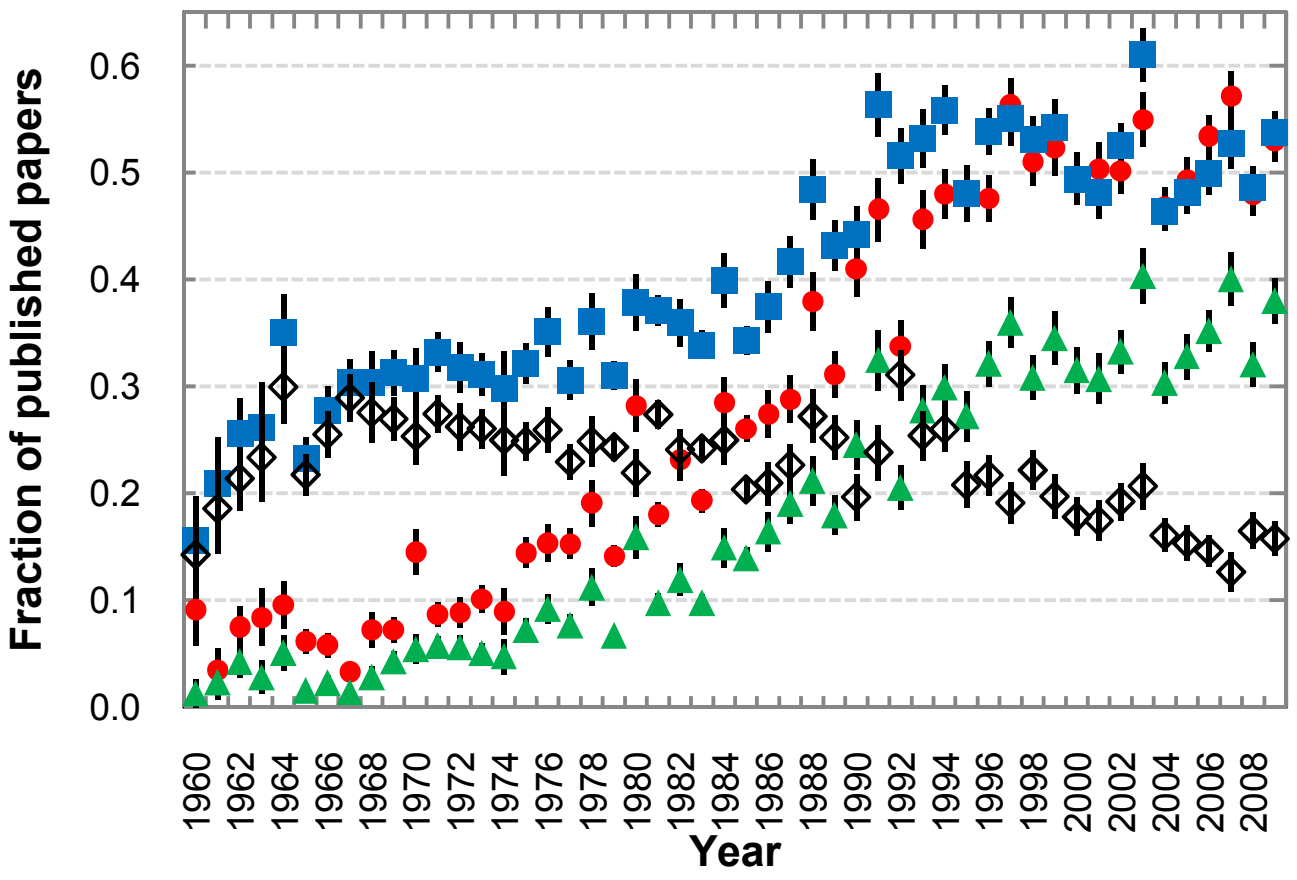

Fig. 1. Fraction of papers published in TNS, mentioning string patterns related to modeling and simulation in the body of the article, as a function of time: any word pattern containing "model" as a substring (blue squares), either "Monte Carlo" or "simulation" (red circles), the Boolean AND of the two previous patterns (green triangles), and the "model" substring without mentioning "Monte Carlo" or "simulation" (white diamonds).

---

T. Basaglia is with CERN, CH-1211 Geneva 23, Switzerland.

Z. W. Bell is the Oak Ridge National Laboratory, Oak Ridge TN 37830; Oak Ridge National Laboratory is managed by UT-Battelle, LLC, for the U.S. Department of Energy under contract DE-AC05-00OR22725.

P. V. Dressendorfer is a consultant and Editor-in-Chief, IEEE TNS.

M. G. Pia is with INFN Sezione di Genova, (telephone: +39 010 3536328, e-mail: MariaGrazia.Pia@ge.infn.it)..

The occurrence of "Monte Carlo" or "simulation" shows a similar trend in TNS and NIM, as illustrated in Fig. 2; nevertheless, these word patterns appear more frequently in TNS than in NIM in the past 20 years, while the trends appear almost identical between 1960 and 1990.

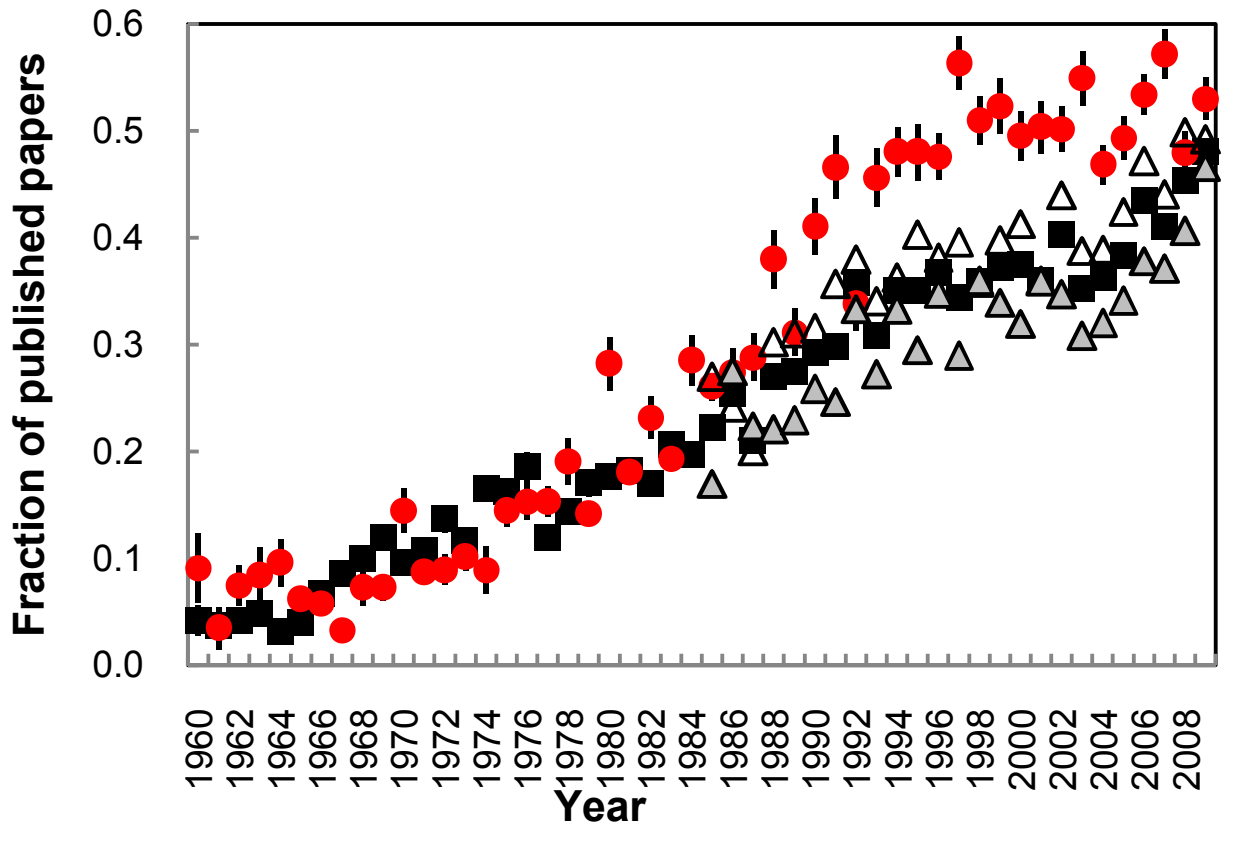

Fig. 2. Fraction of papers published in representative nuclear technology journals, which mention "Monte Carlo" or "simulation": TNS (red circles), NIM A (white triangles), NIM B (grey triangles) and NIM (black squares). The NIM data show the total fraction summed over NIM A and B after the original NIM journal split into two independent journals.

On a global scale one can observe an overall increase of the occurrence of "Monte Carlo" or "simulation" in other related fields, such as medical physics (Fig. 3) and fundamental physics (Fig. 4), nevertheless, in contrast to the trend in TNS and NIM, the pattern of occurrences shows large oscillations.

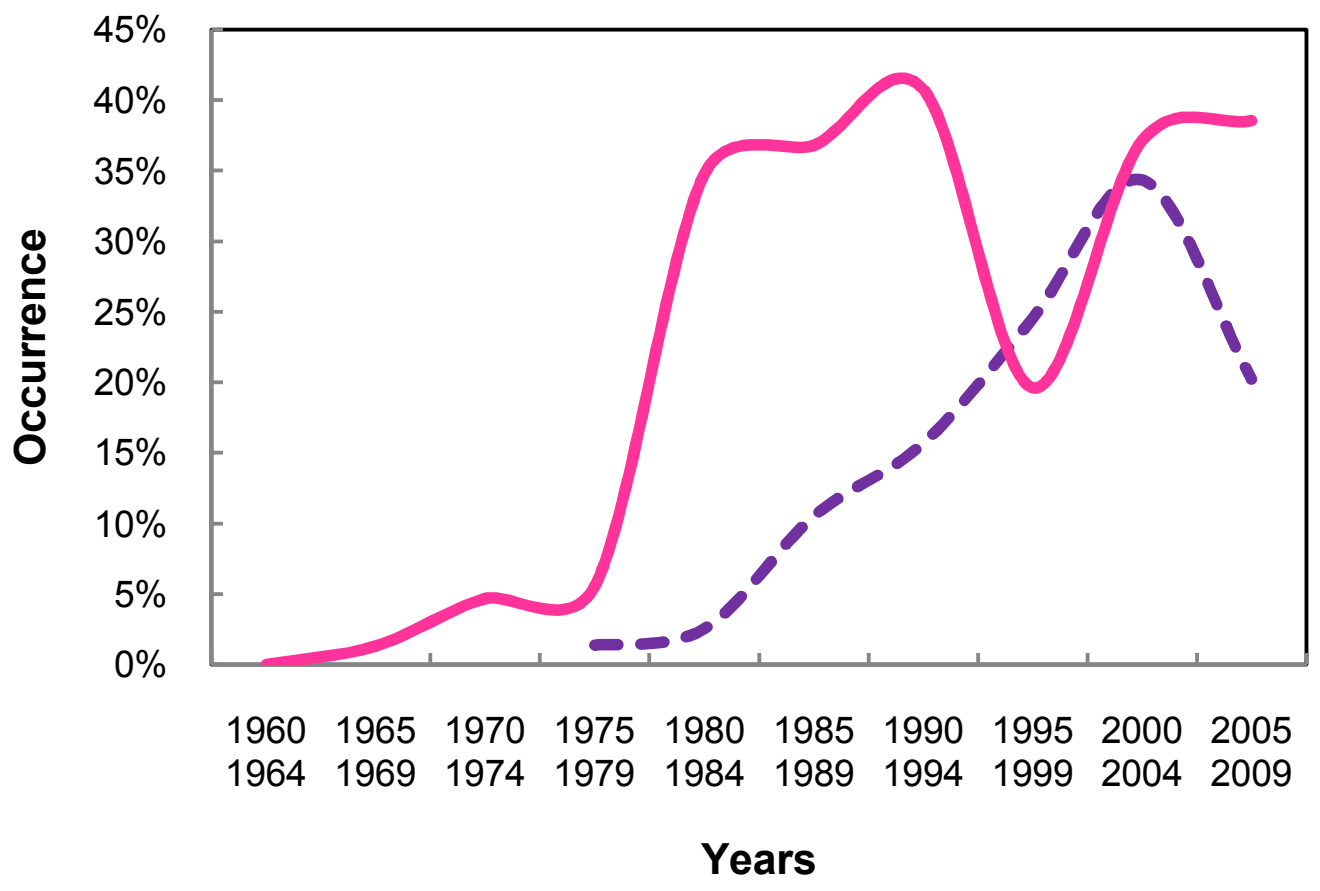

Fig. 3. Fraction of papers published in representative medical physics journals, which mention "Monte Carlo" or "simulation": in Medical Physics (dashed violet line) and in Physics in Medicine and Biology (solid magenta line). Each time bin spans five years.

These patterns show some connection with the time when new, general purpose Monte Carlo systems became available, which were tailored to the requirements of specific experimental fields. GEANT [2] and Geant4 [3][4], which were born from the high energy physics community, were first released respectively in 1978 and at the end of 1998. EGS [5][6], whose simulation capabilities are especially relevant to medical physics applications, was first released in 1978, while MCNPX [7], also widely used in medical physics, became available to the public in 1999, approximately at the same time as Geant4. MCNP [8], which also provides capabilities for the simulation of photon interactions, was first released in 1977.

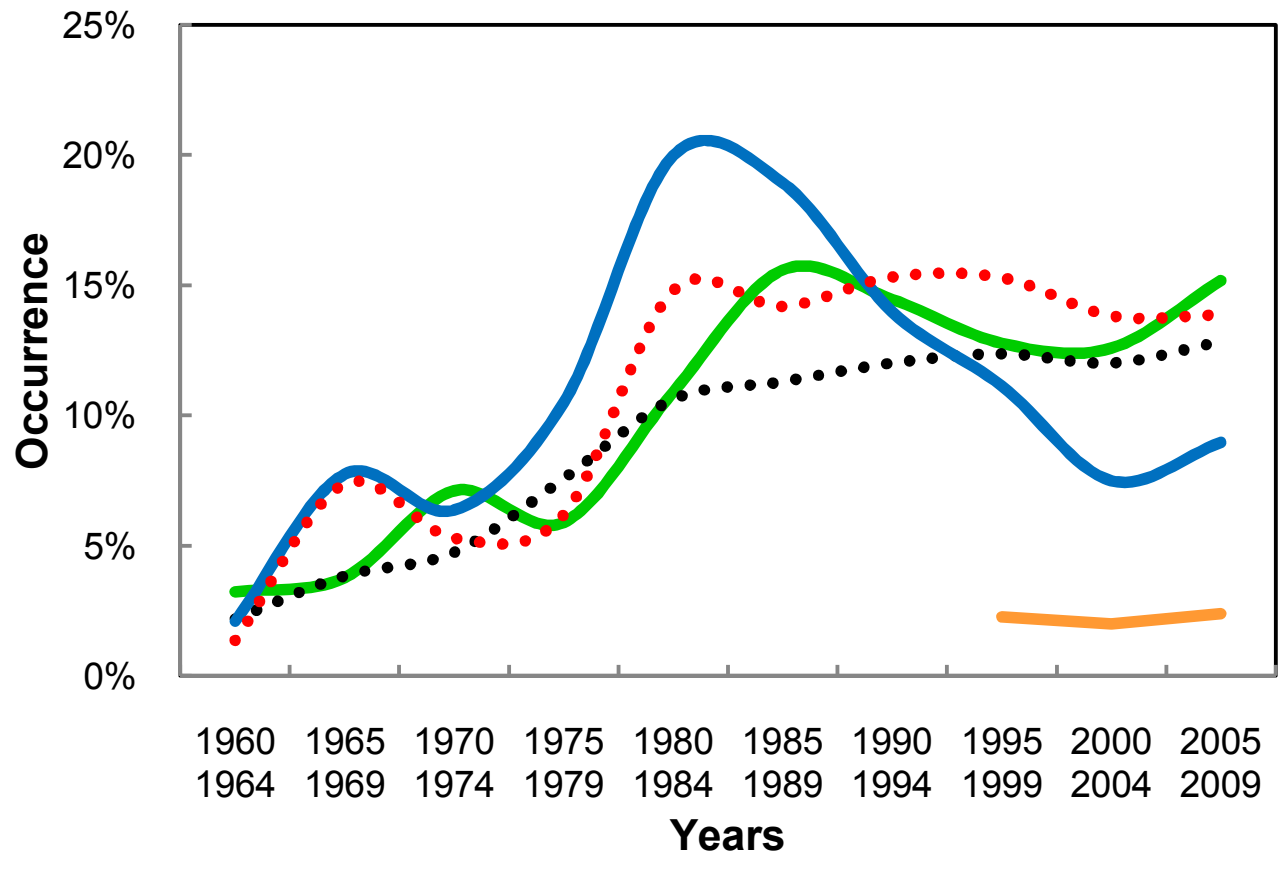

Fig. 4. Fraction of papers published in representative fundamental physics journals, which mention "Monte Carlo" or "simulation": in Physical Review D (solid green line), Physical Review Letters (dotted black line), Nuclear Physics B (solid blue line), Physics Letters B (dotted red line) and the Astrophysical Journal (solid orange line). Each time bin spans five years.

## III. Main Features of Representative Nuclear Technology Journals

The total number of articles published annually by TNS and NIM is shown in Fig. 5.

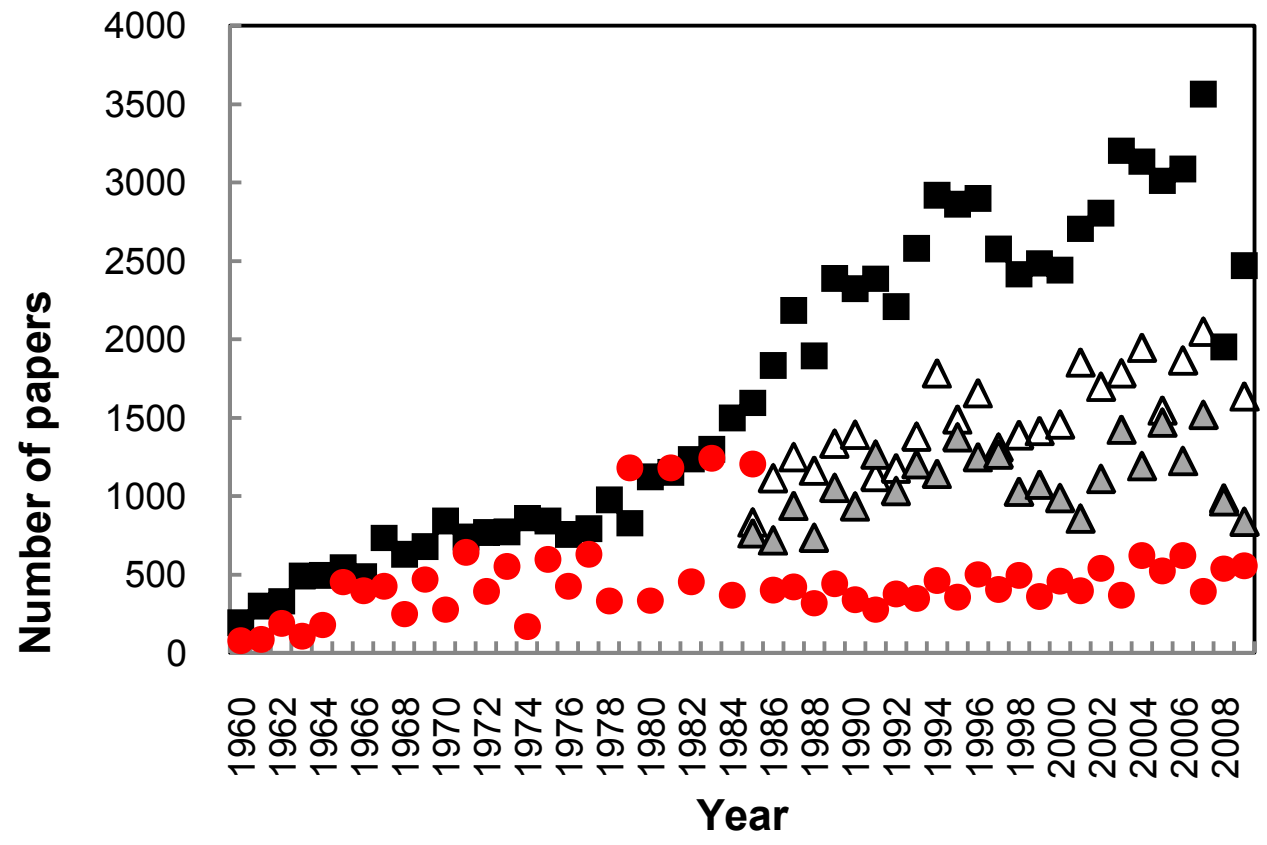

Fig. 5. Total number of papers published each year by TNS (red circles), NIM A (white triangles), NIM B (grey triangles) and NIM (black squares). The NIM data show the total fraction summed over NIM A and B after the original NIM journal split into two independent journals.

More than half of the papers published in TNS in 1990-2009 include a US institute among the authors; in terms of counties, the next major contributors are Italy and France, respectively contributing to 11.4% and 9.9% of the published articles. In terms of institutes, the largest contributor is INFN (6.5%), followed by CERN (4.8%) and the US Naval Research Laboratory (3.9%).

The US also led the list of countries contributing to NIM publications, appearing in 24.2% of the papers; they are followed by Germany (15.4%) and Japan (13.1%). INFN and CERN lead the list of contributing institutes, respectively contributing to 6% and 4.5% of the publications in 1990-2009.

The statistical data reported above derive from Thomson-Reuters' ISI web of Science [9]; the time interval corresponds to the coverage of the subscription available to the authors at the time when the statistics were collected.

## IV. Analysis of Socio-Economic Factors

A number of factors were investigated to verify if their patterns as a function of time showed any correlation with the trend of increased presence of Monte Carlo and simulation in nuclear instrumentation journals:

- expenditures and employment in R&D (research and development),
- academic degrees in science and engineering, and other disciplines,
- outreach in the media,
- financial parameters,
- average income,
- cost of computing equipment,
- the size of experimental collaborations.

In most cases a normalization criterion was applied to the data to allow the quantitative comparison of different observables. The data pertaining to each year – publication fractions or any of the above listed variables – were normalized to the integral over the period spanned by each analysis.

The statistical analysis included the calculation of Pearson's correlation factor and goodness-of-fit tests; the latter exploited the Statistical Toolkit [10][11]. The significance level for the rejection of the null hypothesis of compatibility of the distributions subject to these tests was set to 5%, unless differently specified in the following sections. Pearson's correlation factor is applicable to sample expressed in different units and scales, but it is limited to describing linear relationships and is sensitive to outliers and heavyweight tails; non parametric goodness of fit tests are more generally applicable, but they require manipulations of the original data to produce comparable samples.

### *A. Expenditures and employment in R&D*

The normalized expenditures in R&D of representative countries and organizations [12] are shown in Fig. 6 and Fig. 7 along with the normalized fraction of TNS papers mentioning "Monte Carlo" or "simulation".

Goodness-of-fit tests (Kolmogorov-Smirnov, Anderson-Darling and Cramer-von Mises) show that the expenditure distributions in Fig. 6 and Fig. 7 are all compatible with the TNS distribution with 5% significance, with the exception of UK's and China's expenditures.

According to the same goodness-of-fit tests, TNS "Monte Carlo" or "simulation" fraction is compatible with the distribution of 1983-2007 US employment in science and engineering [12] with 5% significance. The same result holds for the distribution of occupations in specific fields [13] (physics, mathematics and computing, social sciences, life sciences, technicians and programmers) in 1960-2009; nevertheless, this test involves a small sample size, due to the coarse sampling of these distributions over decades.

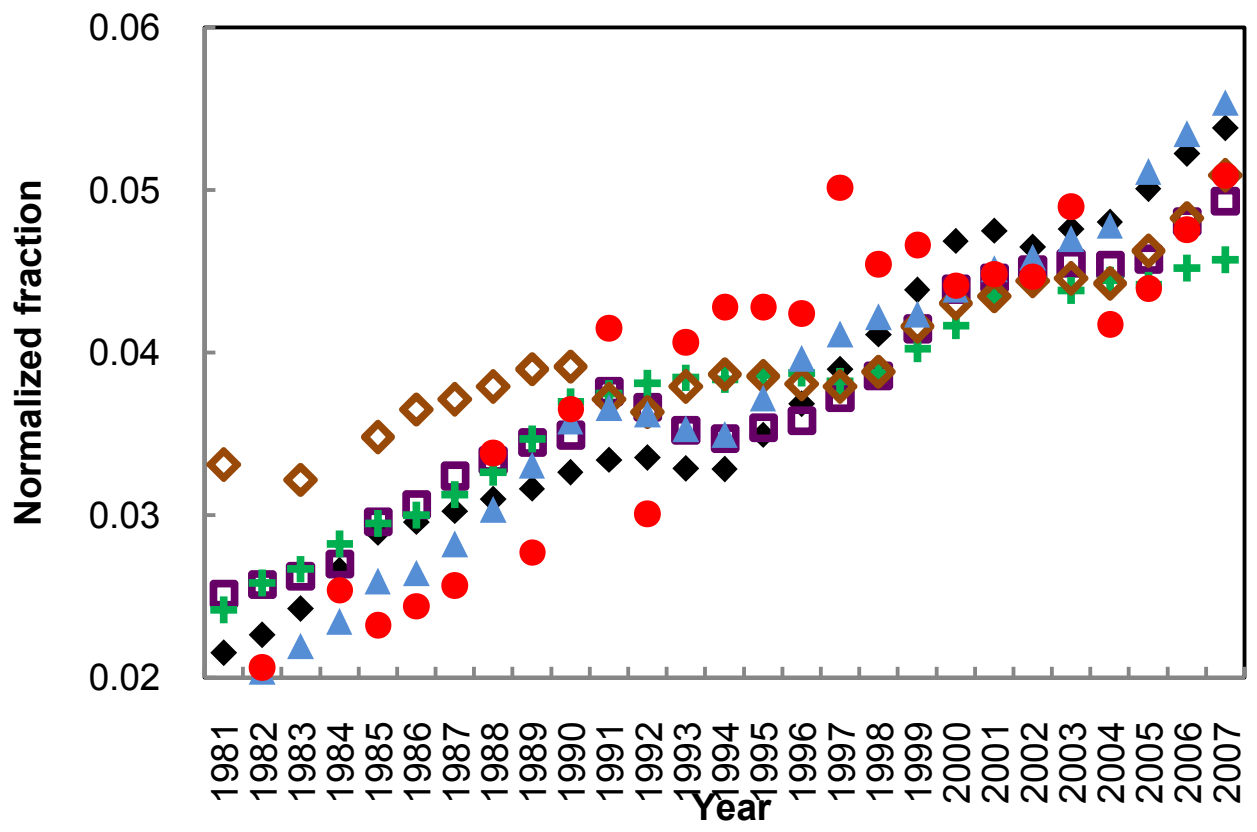

Fig. 6. Gross domestic expenditures on R&D: US (black diamonds), Germany (empty squares), France (green crosses), UK (empty diamonds) and Japan (blue triangles), compared to the normalized fraction of TNS papers mentioning "Monte Carlo" or "simulation" (red circles).

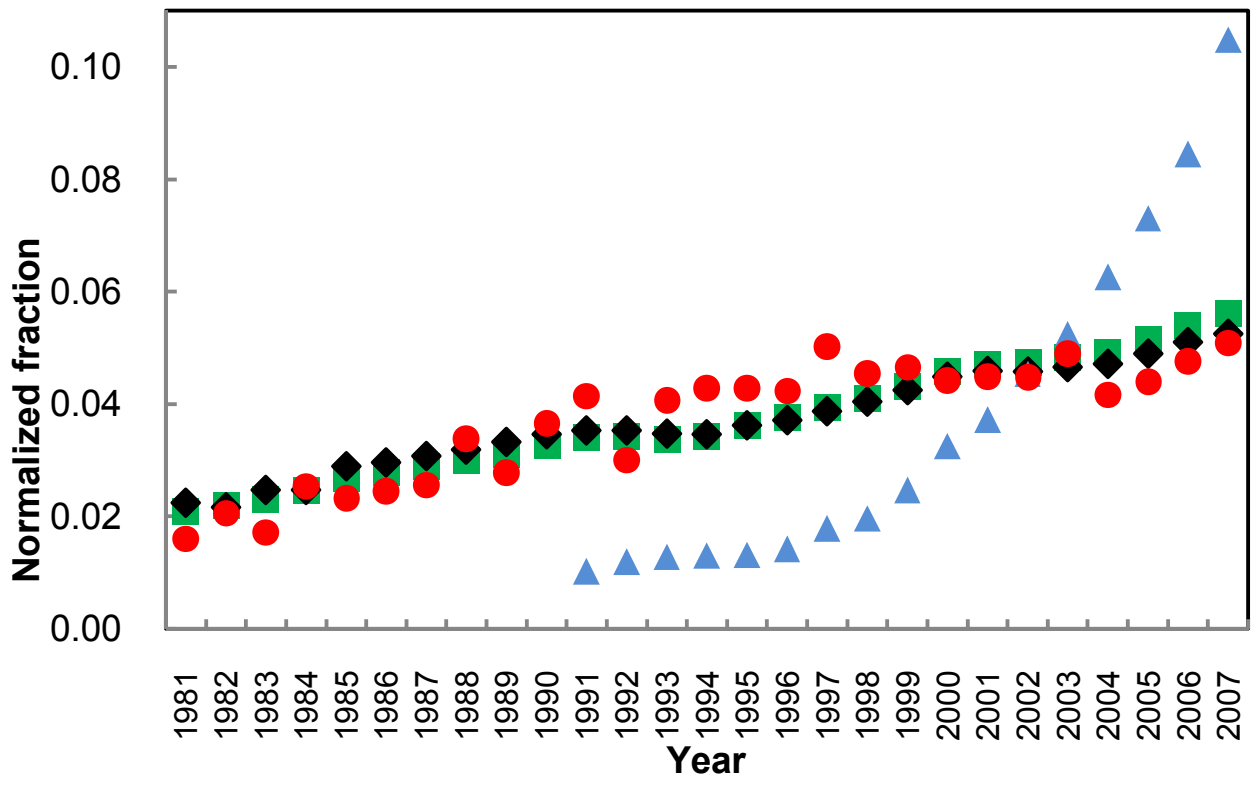

Fig. 7. Gross domestic expenditures on R&D: G7 (black diamonds), OECD (green squares) and China (blue triangles), compared to the normalized fraction of TNS papers mentioning "Monte Carlo" or "simulation" (red circles).

### *B. Academic degrees*

The investigation concerned bachelor, master's and PhD degrees [14][15]; the analysis considered the geographical distribution and disciplines of the degrees.

The distributions of US bachelor degrees in natural sciences, engineering and social sciences are shown in Fig. 8. The outcome of the aforementioned goodness-of-fit tests is controversial: the Kolmogorov-Smirnov test returns p-value 0.502, the Anderson-Darling test rejects the hypothesis of compatibility with TNS "Monte Carlo" or "simulation" fraction distribution with 5% significance, while the Cramer-von Mises test returns p-values between 0.025 and 0.054 associated with the three disciplines. Tests whose statistic involves quadratic terms, like Anderson-Darling and Cramer von Mises, are generally considered to be more powerful than the Kolmogorov-Smirnov test, and the Anderson-darling test in particular is more sensitive to fat tails; nevertheless, the properties of goodness of fit tests, and consequently the identification of the optimal algorithm for a given test case, are still object of research. The use of different tests in this paper mitigates the risk of introducing systematic effects in the conclusions drawn from data analysis based on a single test; discrepant results from different tests, or p-values close to the critical region, suggest caution in drawing firm conclusions in those test cases.

All the considered goodness of fit tests are consistent in accepting the null hypothesis of compatibility between TNS "Monte Carlo" or "simulation" fraction distribution and the distributions of Asian bachelor degrees shown in Fig. 9.

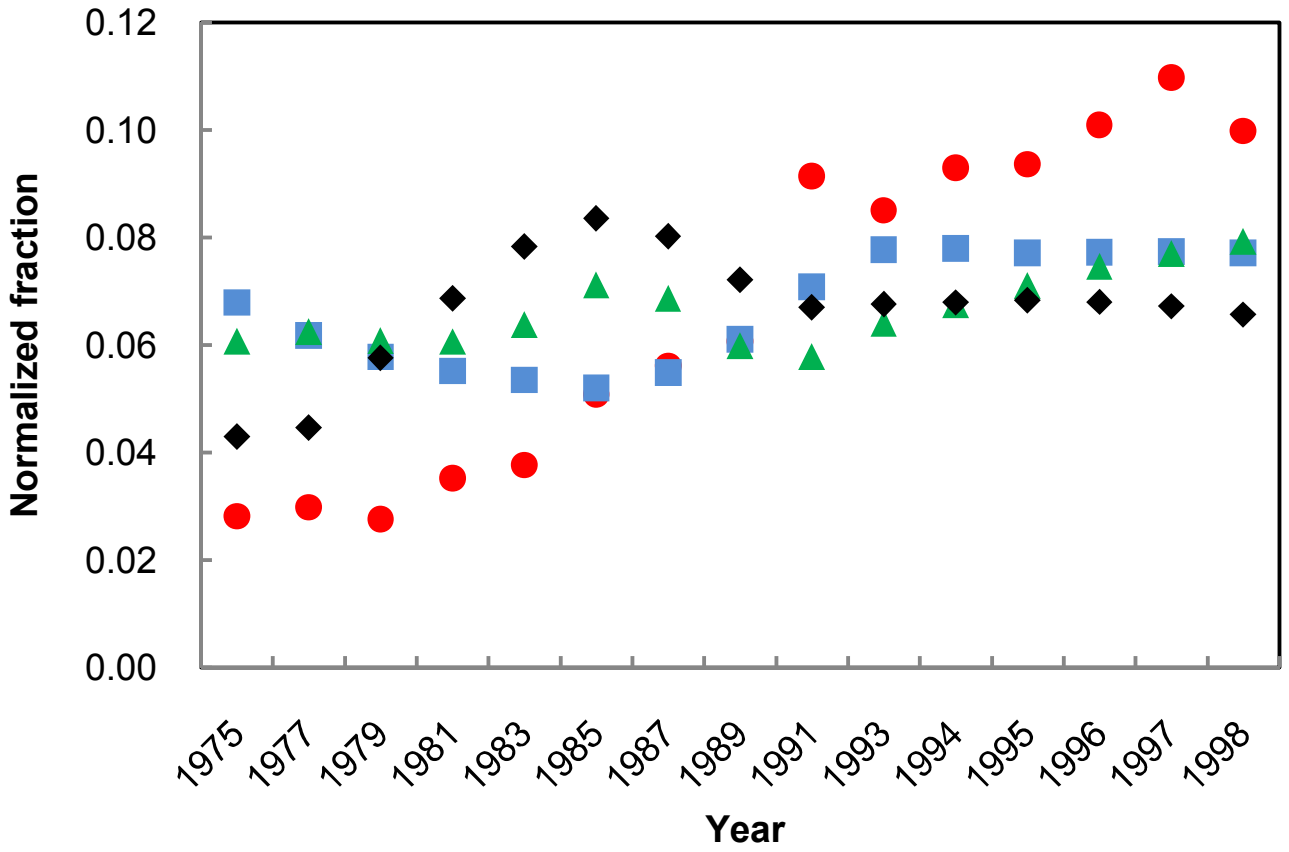

Fig. 8. Normalized number of bachelor degrees in the US: in natural sciences (green triangles), engineering (black diamonds) and social sciences (blue squares), compared to the normalized fraction of TNS papers mentioning "Monte Carlo" or "simulation" (red circles).

The distribution of master's degrees is shown in Fig. 10. The Kolmogorov-Smirnov, Anderson-Darling and Cramer-von Mises confirm the hypothesis of compatibility with the distribution of fraction of TNS papers mentioning "Monte Carlo" or "simulation" for all disciplines, with the exception of computing.

The distributions of doctoral degrees in science and engineering in Europe, Asia and the US are shown in Fig. 11; they concern the years 1975-1999. The distributions for Europe and Asia are compatible with the fraction of TNS papers mentioning "Monte Carlo" or "simulation" with 5% significance; as to the comparison concerning the US distribution, the Anderson-Darling test rejects the hypothesis of compatibility with TNS distribution with 5% significance, while the Kolmogorov-Smirnov and Cramer-von Mises tests result in p-values close to the critical region.

The distributions of natural sciences and engineering doctoral degrees of selected countries are shown in Fig. 12; they the years 1993-2006. The goodness-of-fit tests reject the hypothesis of compatibility with the fraction of TNS papers mentioning "Monte Carlo" or "simulation" only for distributions of doctoral degrees in China and South Korea.

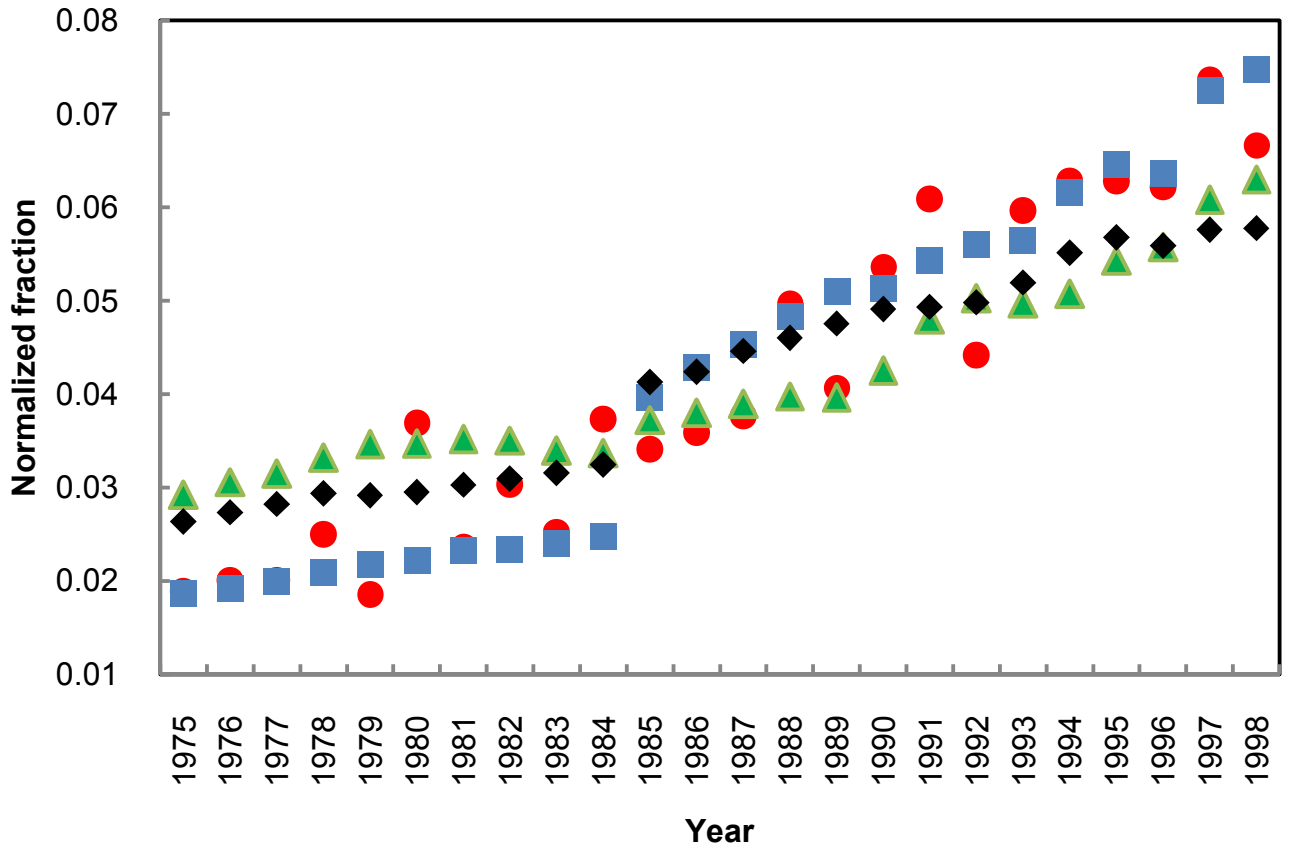

Fig. 9. Normalized number of bachelor degrees in Asia: in natural sciences (green triangles), engineering (black diamonds) and social sciences (blue squares), compared to the normalized fraction of TNS papers mentioning "Monte Carlo" or "simulation" (red circles).

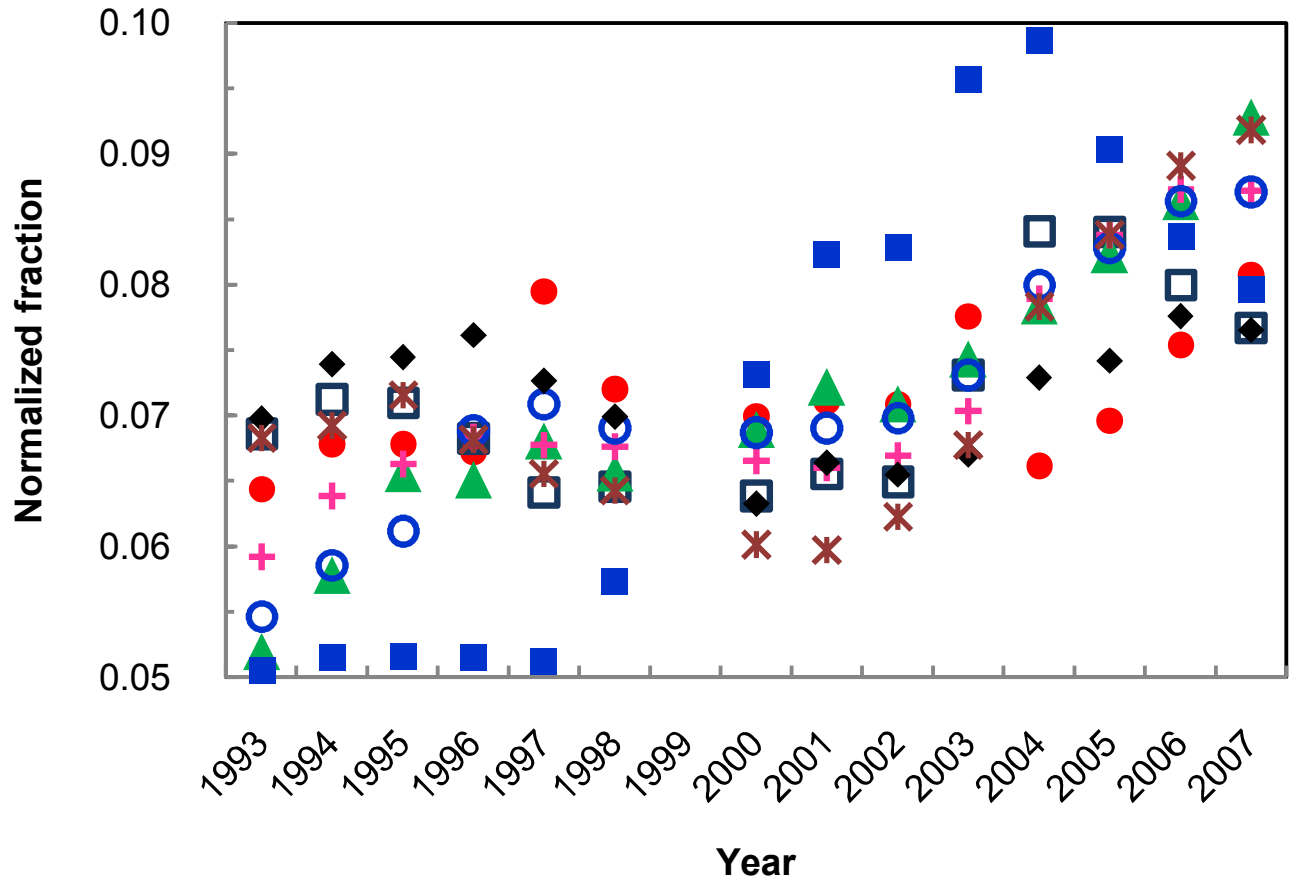

Fig. 10. Normalized number of master's degrees in the US: in physical sciences (black diamonds), psychology (green triangles), engineering (empty squares), social sciences (pink crosses), computer sciences (blue squares), biological and agricultural sciences (empty circles) and mathematics (brown asterisks), compared to the normalized fraction of TNS papers mentioning "Monte Carlo" or "simulation" (red circles).

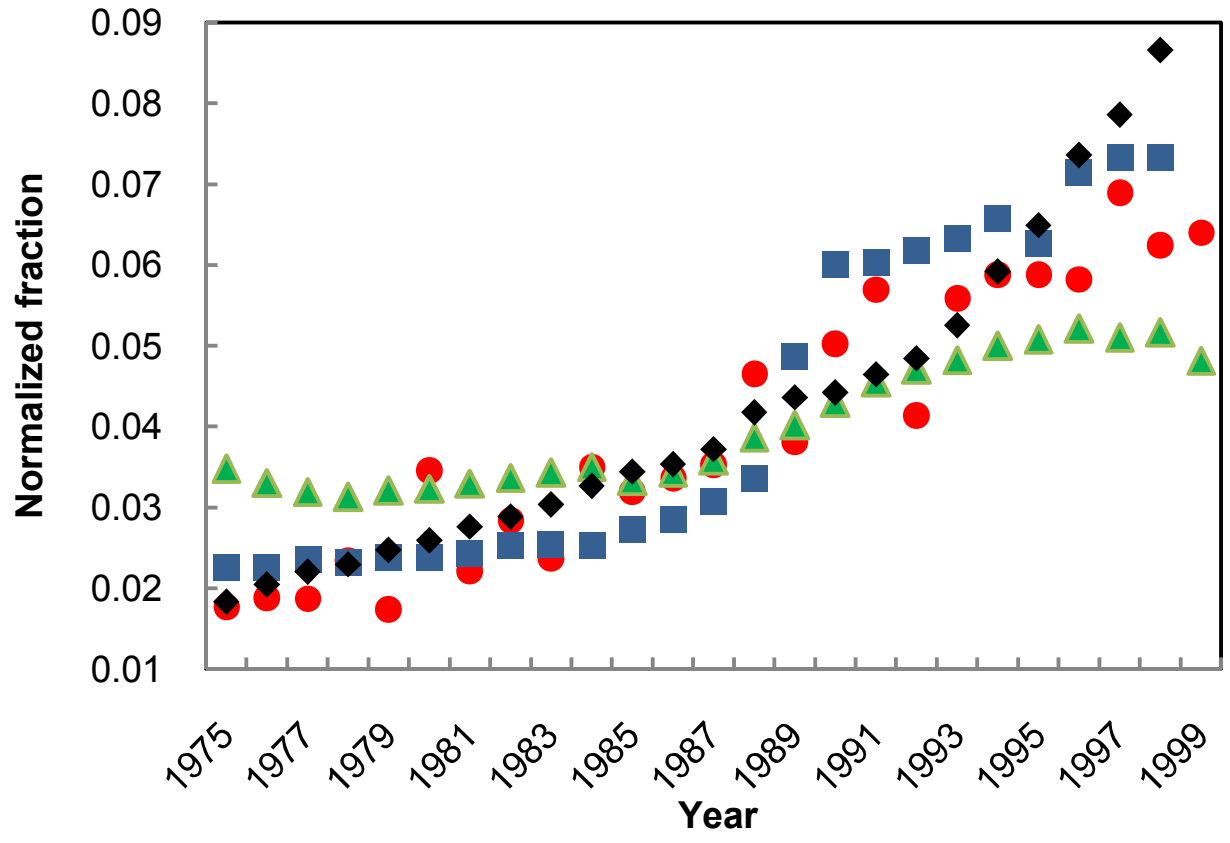

Fig. 11. Normalized number of doctoral science and engineering degrees in US (green triangles), Europe (blue squares) and Asia (black diamonds), compared to the normalized fraction of TNS papers mentioning "Monte Carlo" or "simulation" (red circles).

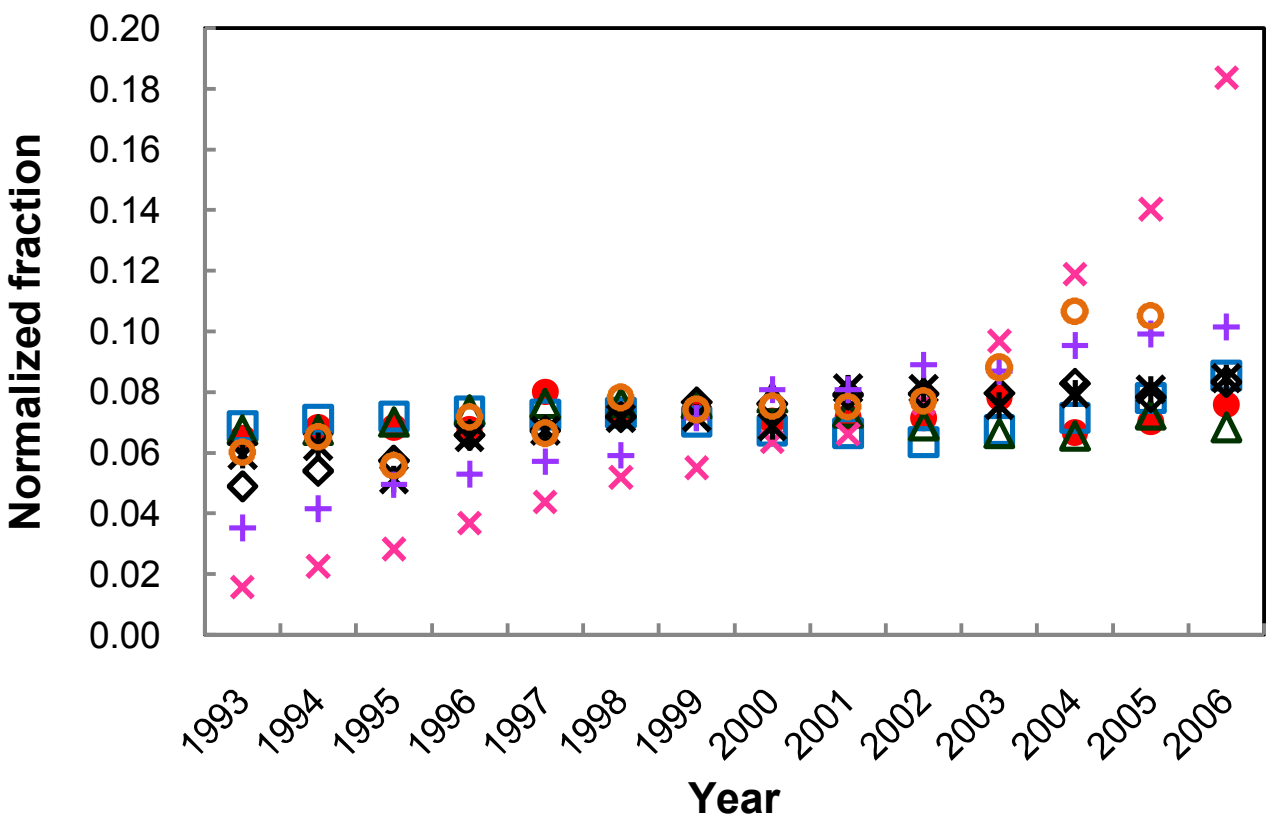

Fig. 12. Normalized number of natural sciences and engineering PhDs in US (empty squares), Germany (empty triangles), China (x), UK (asterisks), Japan (empty diamonds), South Korea (crosses) and India (empty circles), compared to the normalized fraction of TNS papers mentioning "Monte Carlo" or "simulation" (red circles).

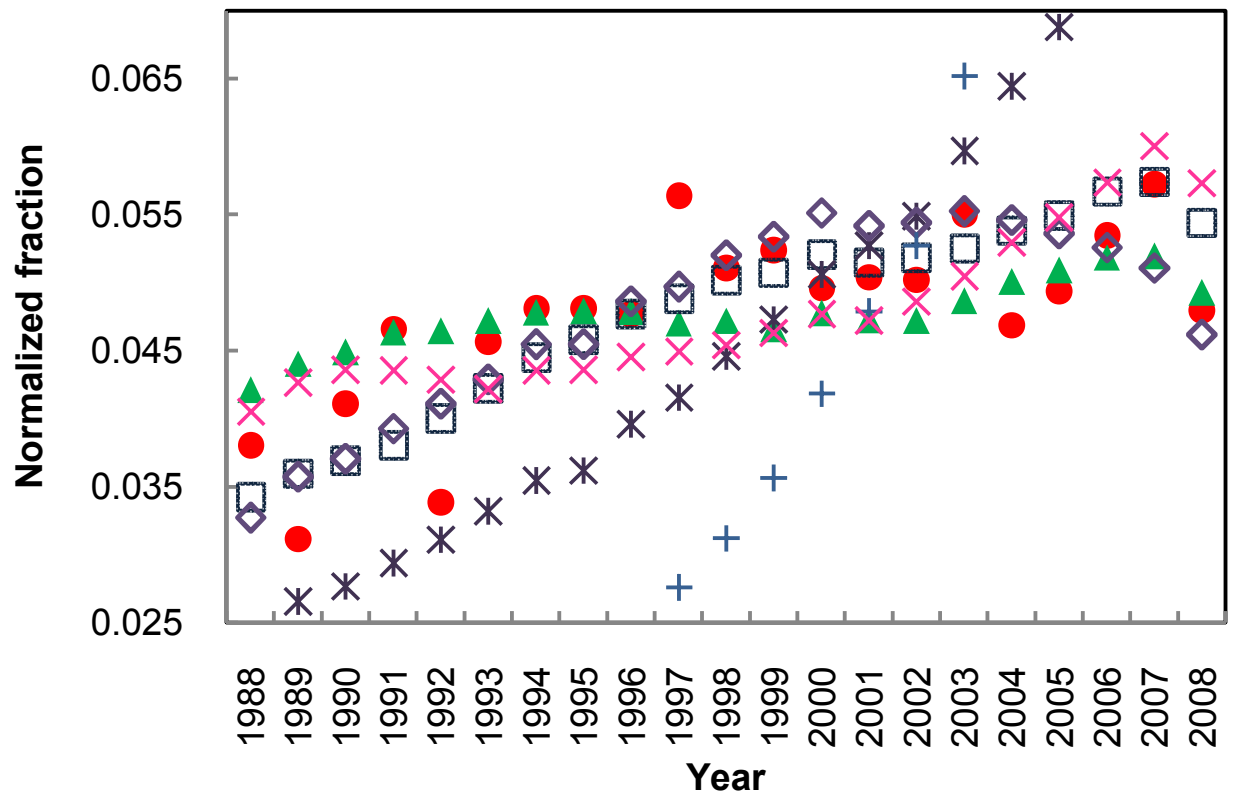

Fig. 13. Normalized number of publications in selected geographical areas: US (green triangles), Europe (empty squares), Asia (asterisks), Japan (empty diamonds), China (crosses) and rest of the world (pink x), compared to the normalized fraction of TNS papers mentioning "Monte Carlo" or "simulation" (red circles).

A further analysis examined the correlation of distributions of US doctoral degrees across various disciplines in the years 1993-2007. Only the trend of bio-medical PhDs was found incompatible with the fraction of TNS papers mentioning "Monte Carlo" or "simulation"; the trends of engineering, physics, mathematics, computing, agriculture and social sciences doctoral degrees are all consistent with the TNS distribution.

### *C. Publications*

The normalized distributions of science and engineering articles published in scholarly journals are shown in Fig. 13 for selected geographical areas [16].

The hypothesis of compatibility with the fraction of TNS papers mentioning "Monte Carlo" or "simulation" is rejected for the distribution associated with China; regarding the distribution of Asian publications, the Anderson-Darling test rejects the hypothesis of compatibility, while the Kolmogorov-Smirnov and Cramer-von Mises tests do not.

### *D. Outreach*

The possible influence by the media has been investigated. A plot illustrating the coverage of selected science topics by major US networks is shown in Fig. 14; it concerns bio-medical research topics and science, space and technology research. The data reflect the annual minutes of story coverage on these topics by ABC, CBS, and NBC, out of approximately 15000 total annual minutes on weekday nightly newscasts. Excluded from science, space, and technology are forensic science and media content. Excluded from biotechnology and basic medical research are stories on clinical research and medical technology.

Regarding the hypothesis of compatibility with the trend of TNS fraction of papers mentioning "Monte Carlo" or "simulation", the Anderson-Darling test rejects it for both topics, the Kolmogorov-Smirnov test accepts it, while the Cramer-von Mises test rejects it for what concerns the bio-medical news coverage and accepts it for the coverage of science, space and technology research.

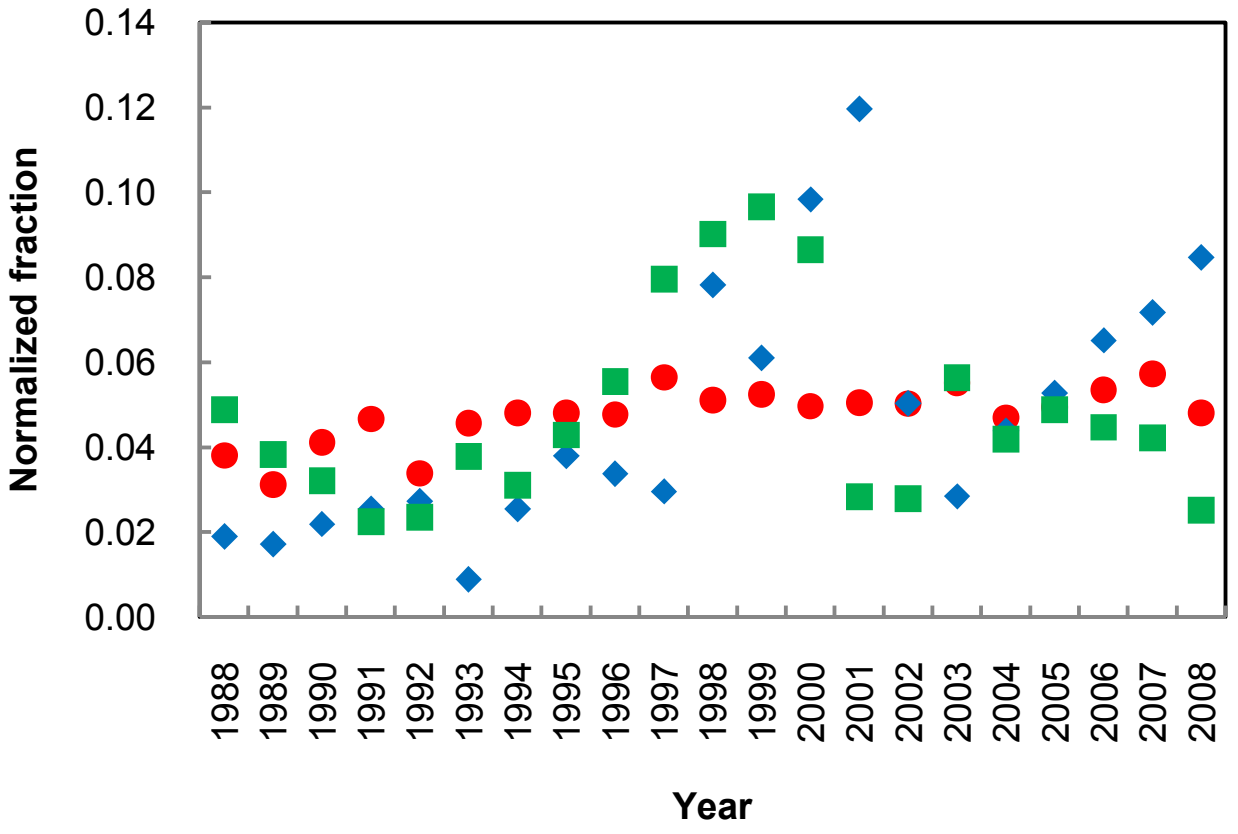

Fig. 14 Normalized annual time of story coverage by major US networks ABC, CBS, and NBC: bio-medical research topics (blue diamonds) and science, space and technology research topics (green squares), compared to the normalized fraction of TNS papers mentioning "Monte Carlo" or "simulation" (red circles).

### *E. Financial indicators*

A set of financial indicators associated with US, European and Japanese stock markets has been analyzed to identify possible correlations with trends in the fraction of TNS papers mentioning "Monte Carlo" or "simulation". The following indices have been evaluated:

- Dow Jones Industrial Average (years 1960-2009)
- Deutscher Aktien IndeX (years 1987-2009)
- FTSE 100 (years 1970-2009)
- Nikkei 225 (years 1970-2009)

No significant correlation is observed with any of these financial indices; an example is shown in Fig. 15. Although there is a general agreement between the trends of the Dow Jones Industrial Average (DJIA) and the fraction of TNS papers mentioning "Monte Carlo" or "simulation", the sharp peaks and valleys of the DJIA are not reproduced in paper statistics. Most notably, the strong bull market from 1996 to 1999 is not reflected in the TNS data, which begins its run-up nearly 20 years prior.

No correlation with gold price is visible in Fig. 16.

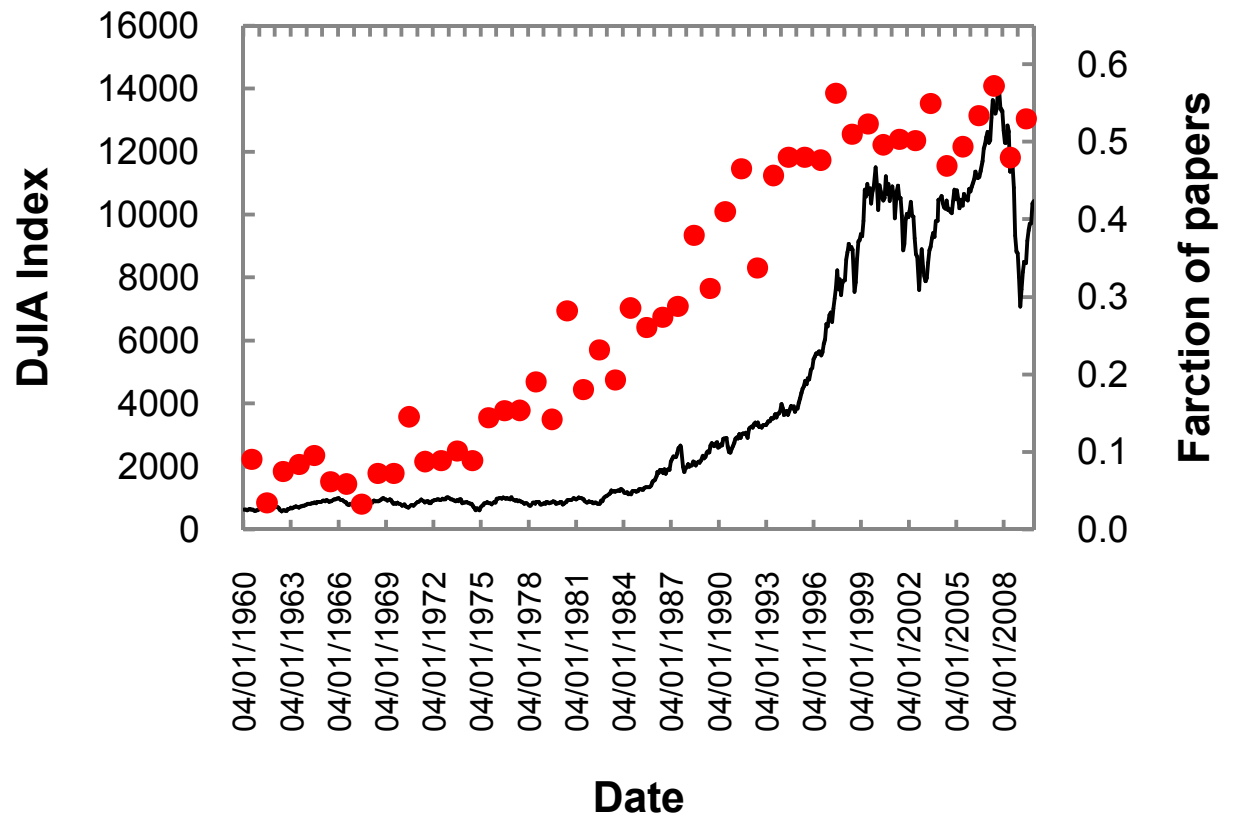

Fig. 15. Dow Jones Industrial Average index (histogram), compared to the fraction of TNS papers mentioning "Monte Carlo" or "simulation" (red circles).

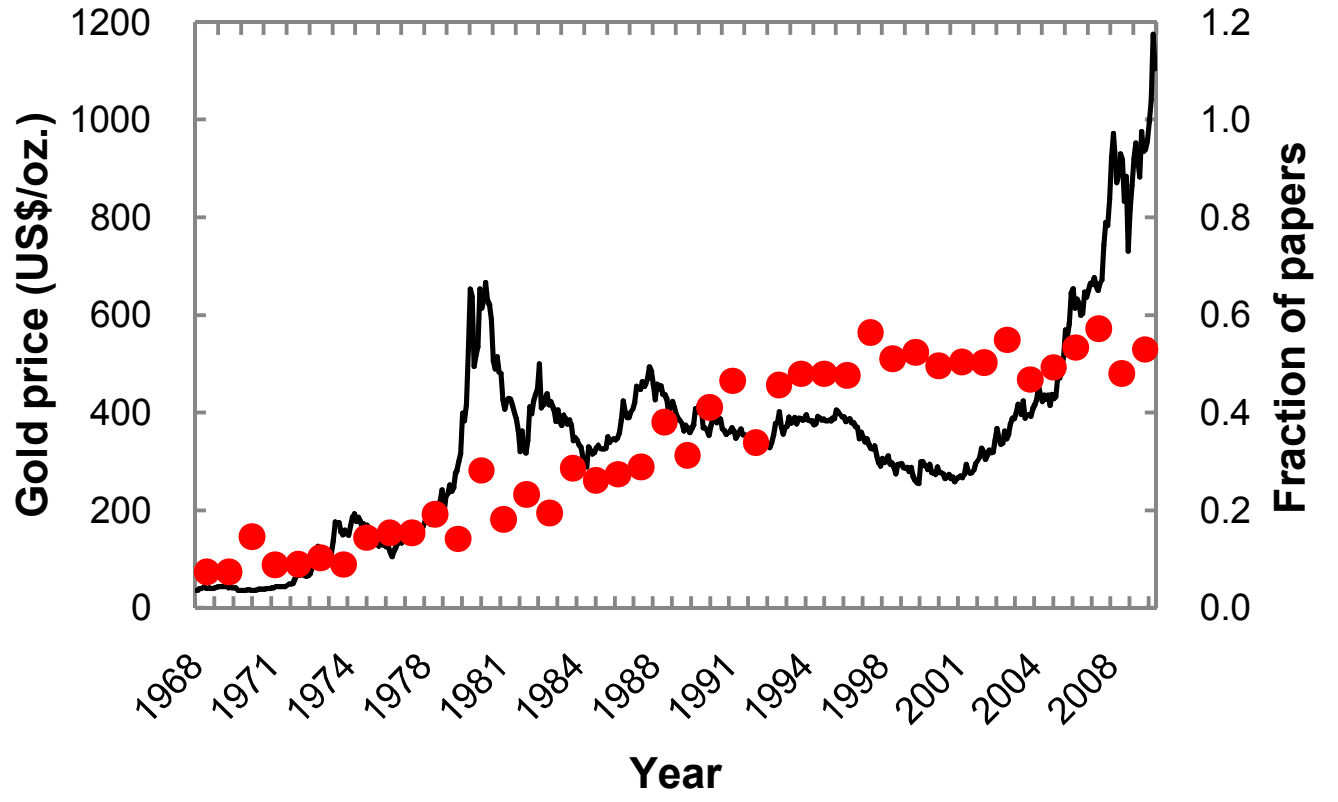

Fig. 16. Gold price (histogram), compared to the fraction of TNS papers mentioning "Monte Carlo" or "simulation" (red circles).

## F. *Income*

Possible effects related to the income *pro capite* in selected countries have been investigated.

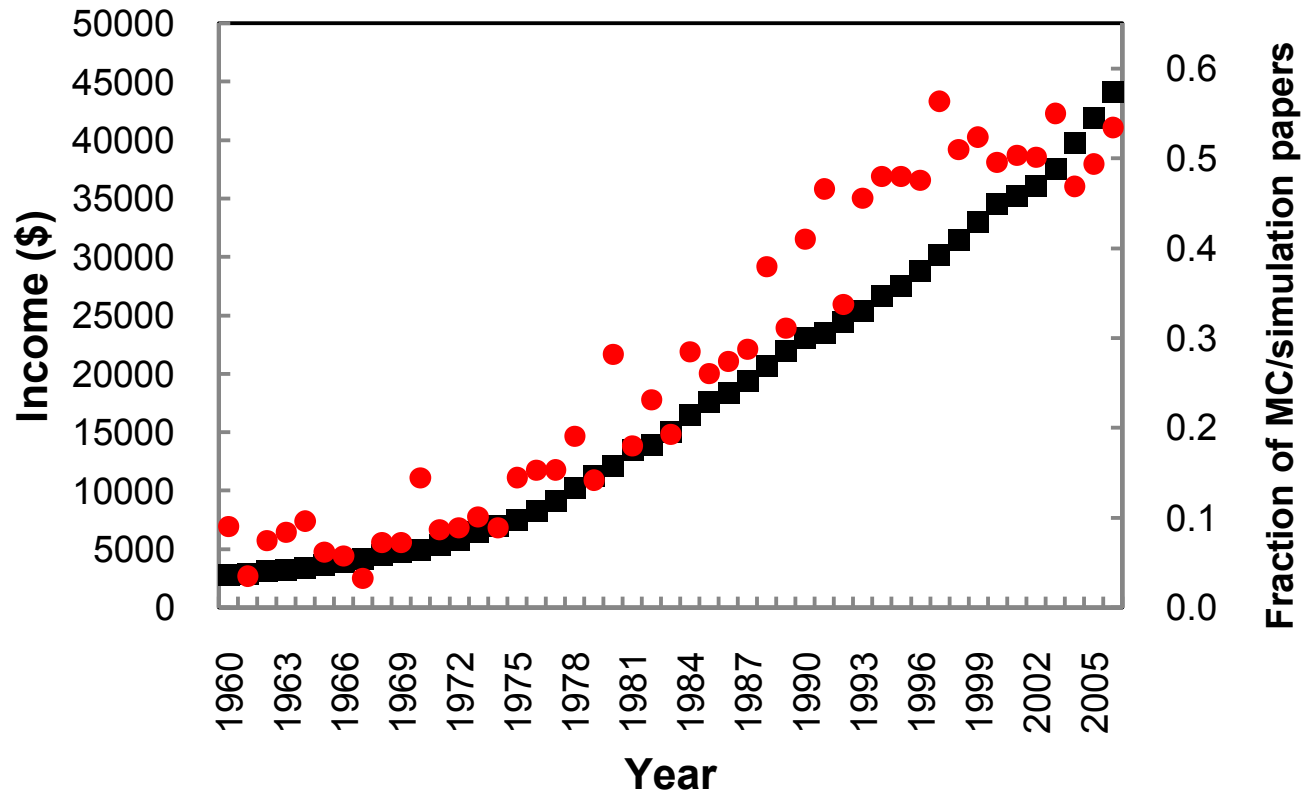

Fig. 17. Average US income (black squares), compared to the fraction of TNS papers mentioning "Monte Carlo" or "simulation" (red circles).

Correlation between the average income and the trend of TNS fraction of papers mentioning "Monte Carlo" or "simulation" is clearly visible in the data associated with so-called first world countries (US, Europe, Japan, Australia), shown in Fig. 17 to Fig. 21; the effect is quantitatively confirmed by the associated Pearson's correlation factor reported in Table I.

Emerging countries like China and Korea exhibit largely different trends with respect to the trend of simulation publications in the past two decades; an example is illustrated in Fig. 22. The income in other countries, like Brazil and Argentina, shows a general trend of compatibility, although subject to large local fluctuations in some periods of economic difficulties; an example is plotted in Fig. 23.

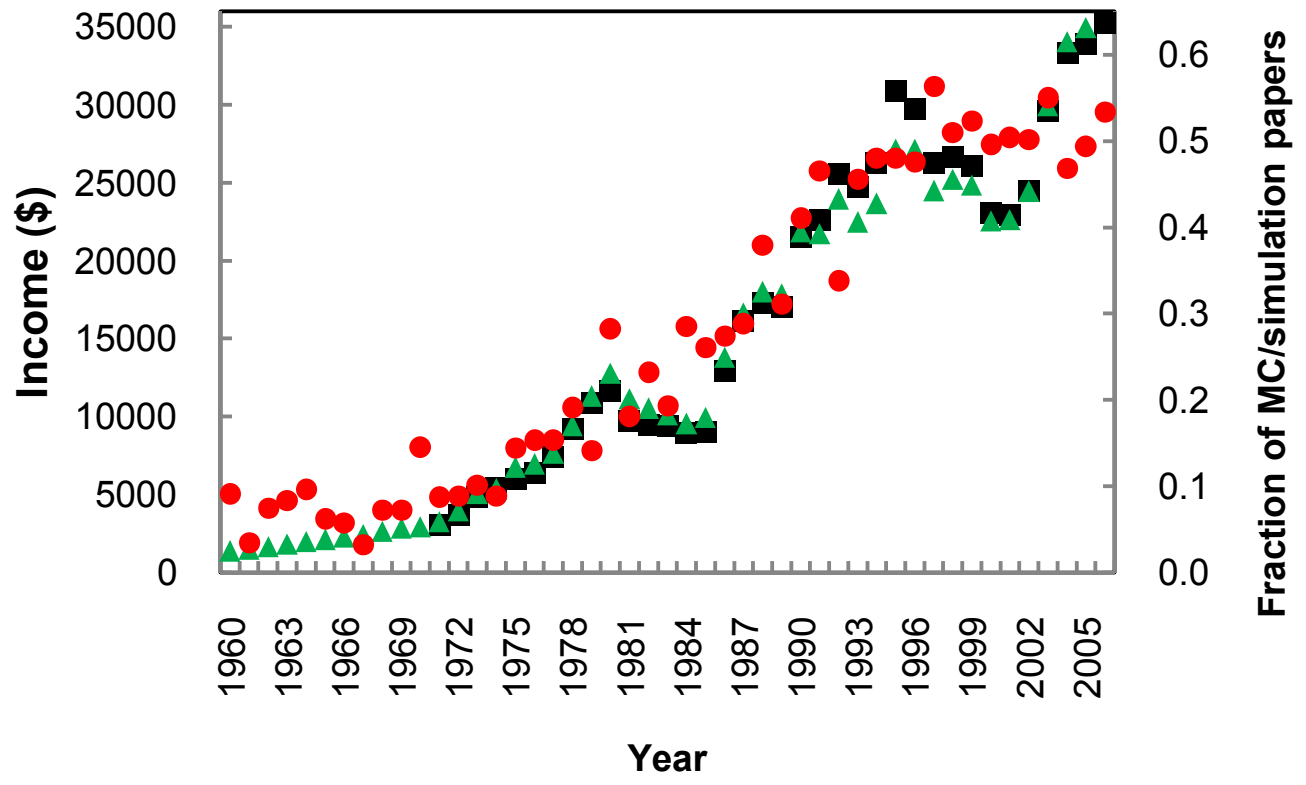

Fig. 18. Average German (black squares) and French (green triangles) income, compared to the fraction of TNS papers mentioning "Monte Carlo" or "simulation" (red circles).

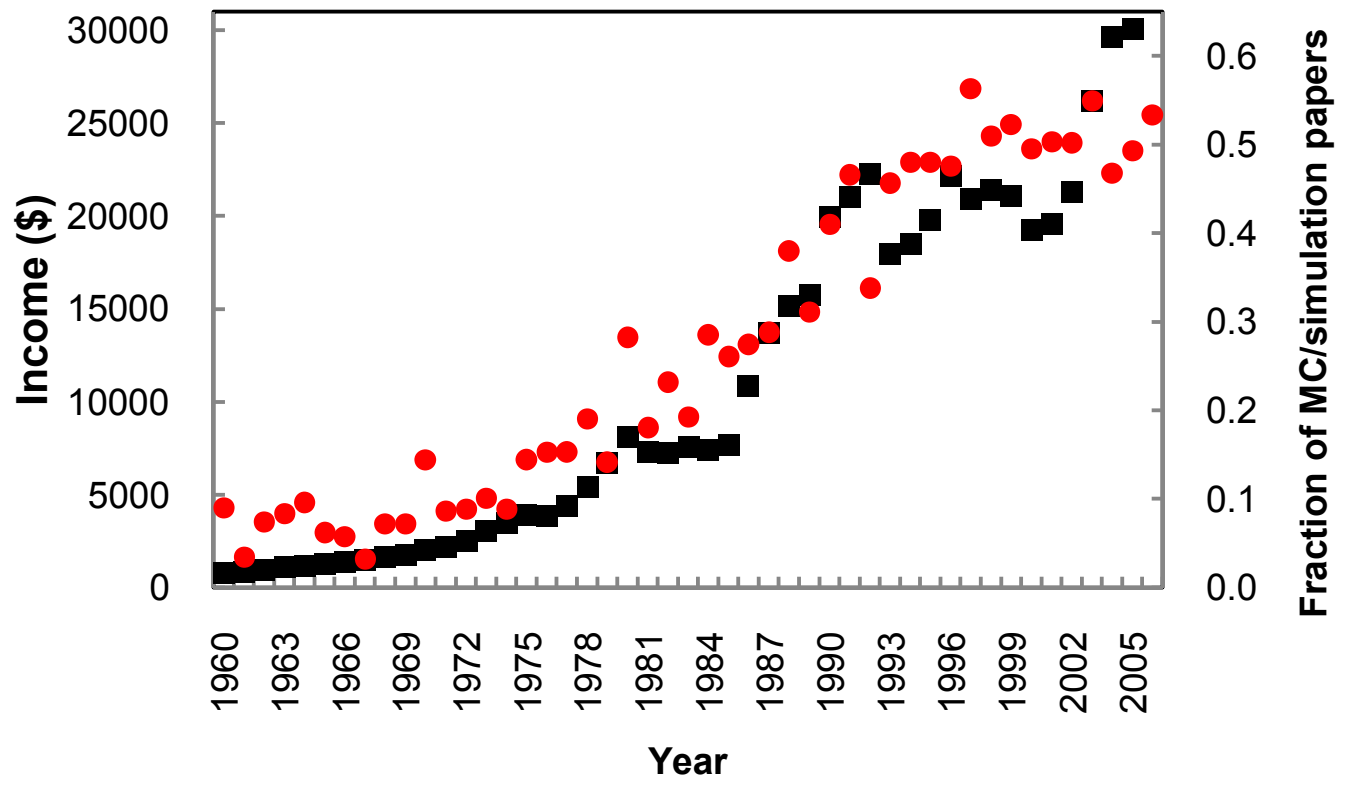

Fig. 19. Average Italian income (black squares), compared to the fraction of TNS papers mentioning "Monte Carlo" or "simulation" (red circles).

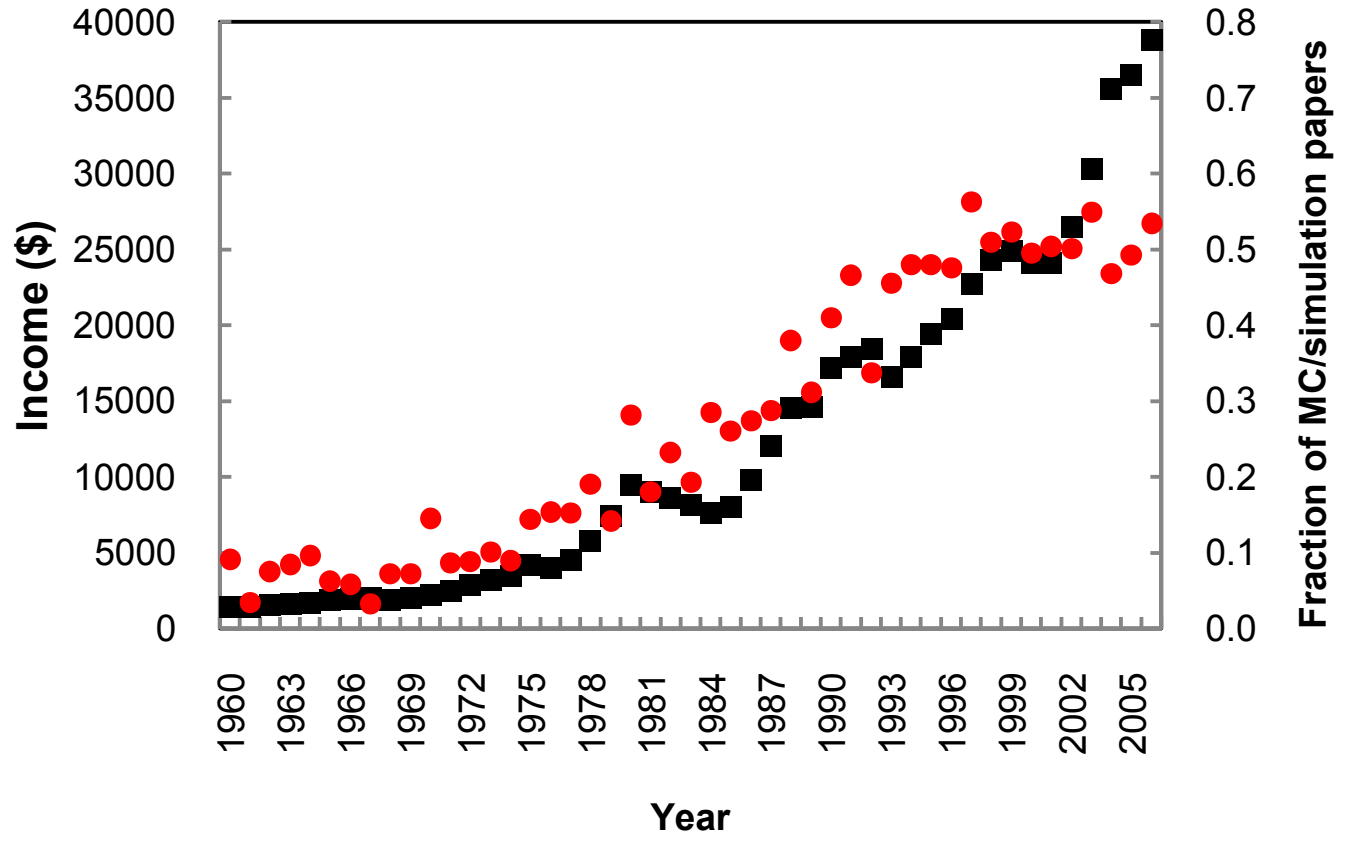

Fig. 20. Average UK income (black squares), compared to the fraction of TNS papers mentioning "Monte Carlo" or "simulation" (red circles).

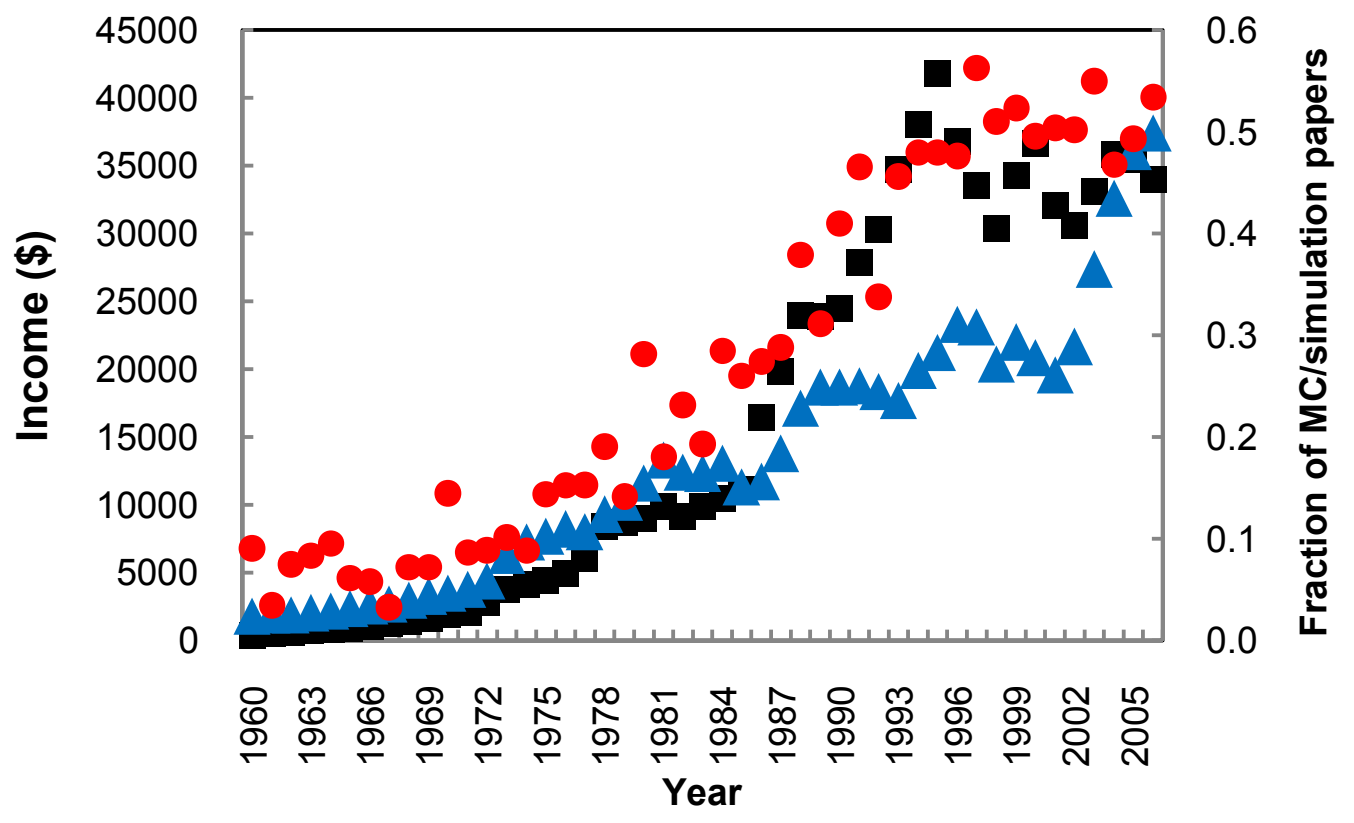

Fig. 21. Average Japanese (black squares) and Australian (blue triangles) income, compared to the fraction of TNS papers mentioning "Monte Carlo" or "simulation" (red circles).

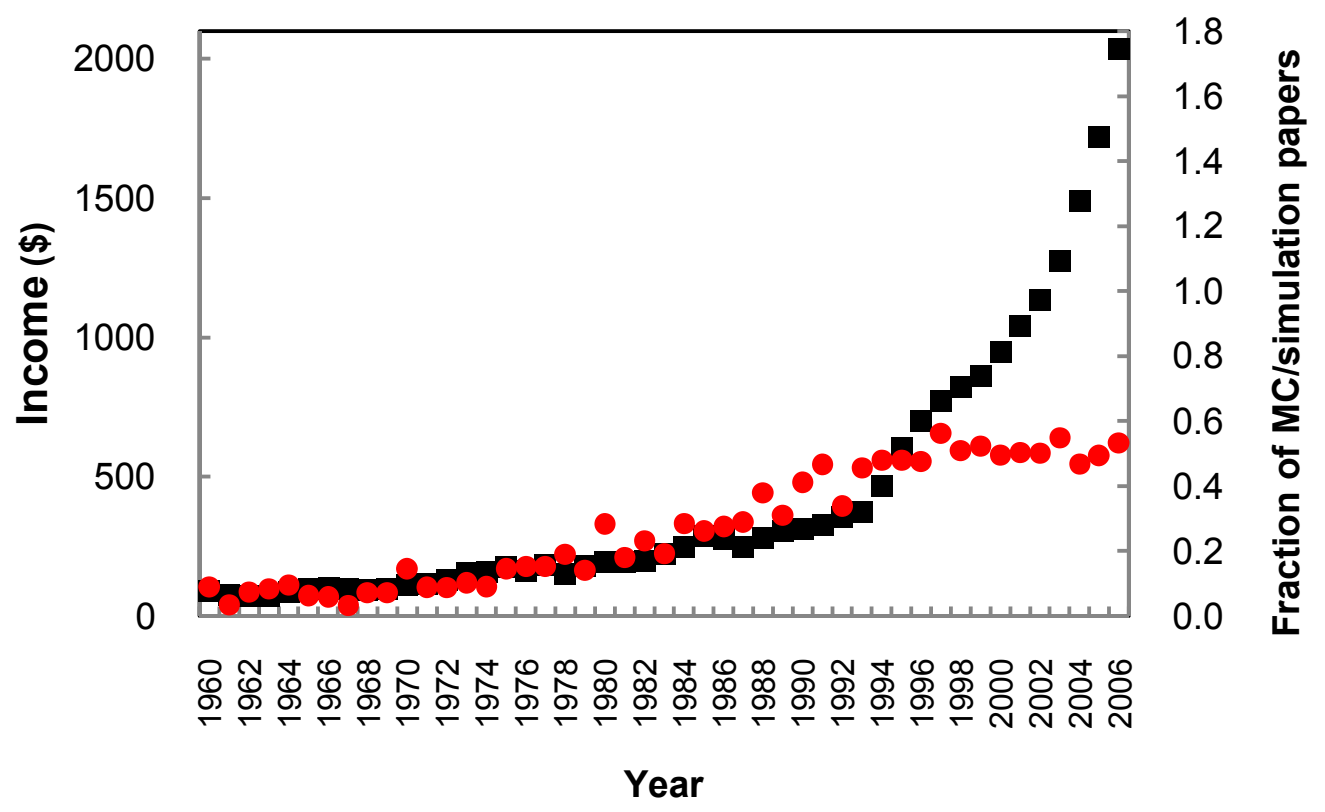

Fig. 22. Average Chinese income (black squares), compared to the fraction of TNS papers mentioning "Monte Carlo" or "simulation" (red circles).

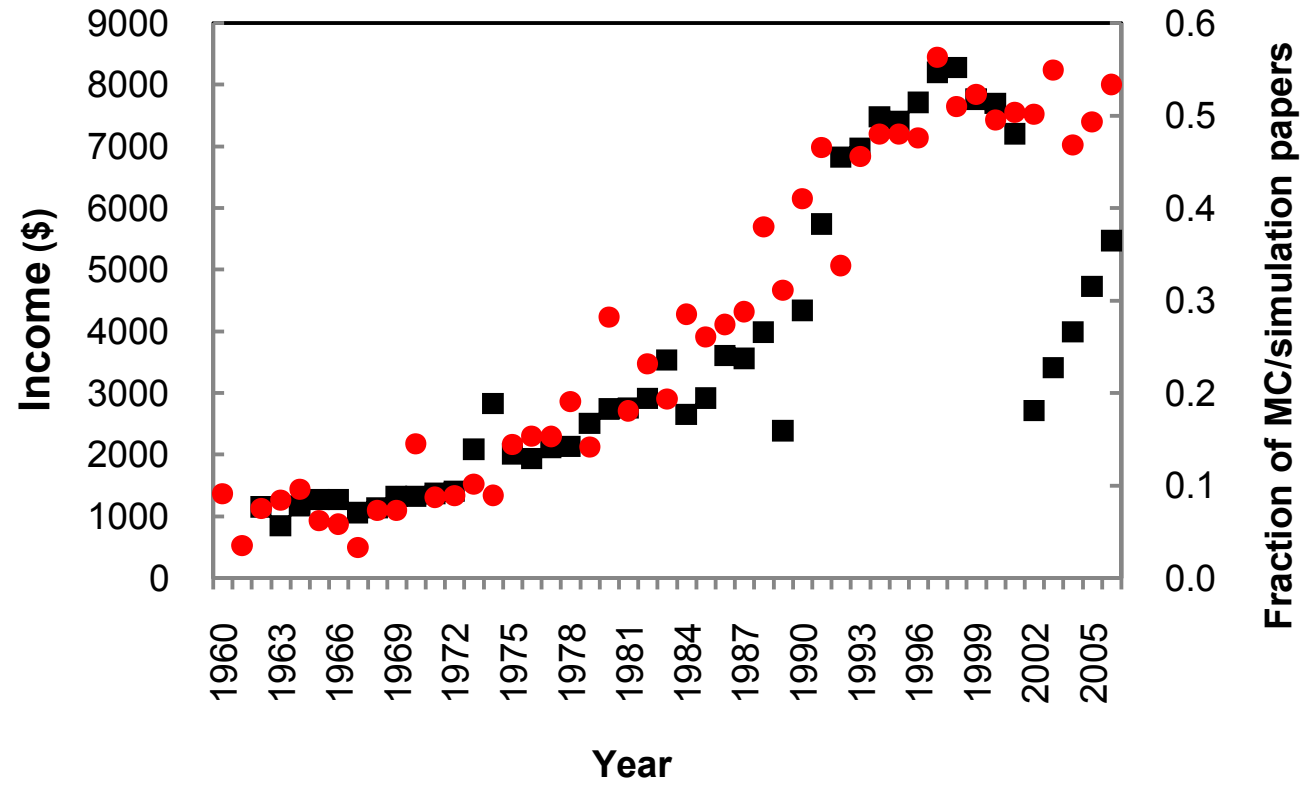

Fig. 23. Average Argentinian income (black squares), compared to the fraction of TNS papers mentioning "Monte Carlo" or "simulation" (red circles).

It is worthwhile to remark that an increase in the average income means an increase in the cost of labor for experimental research.

## G. *Computing costs*

The cost of computing equipment has been decreasing over la past fifty years, while the computing and storage facilities available to researchers have increased dramatically. There is no doubt that these facts have contributed to increase the use of simulation in experimental research. Large scale simulation productions, based on detailed models of experimental set-ups, are easily feasible.

Nevertheless, even if the availability of low-cost computing facilities is certainly a major factor in the current widespread use of simulation in experimental physics, computing costs have been decreasing at a much faster rate than the growth of simulation in nuclear technology publications. An example of the different rates of variation is shown in Fig. 24.

TABLE I. CORRELATION BETWEEN AVERAGE INCOME AND FRACTION OF PAPERS MENTIONING MONTE CARLO OR SIMULATION

| Country | Correlation factor |
|---|---|
| US | 0.96 |
| Germany | 0.93 |
| France | 0.95 |
| UK | 0.94 |
| Italy | 0.95 |
| Japan | 0.97 |
| Australia | 0.94 |

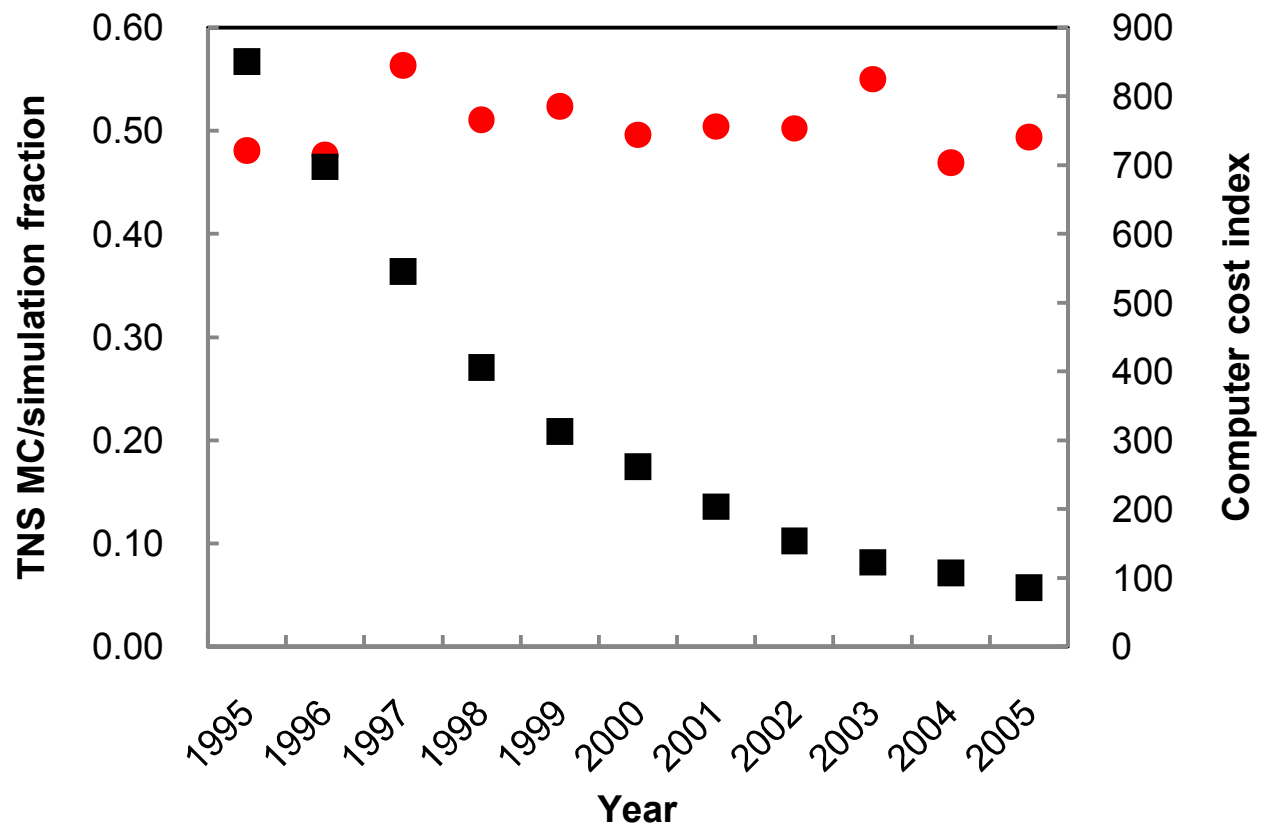

Fig. 24. Computing cost index (black squares) [18], as defined the US Bureau of Labor Statistics, compared to the fraction of TNS papers mentioning "Monte Carlo" or "simulation" (red circles) in the years 1995-2005.

## H. *HEP collaborations*

Nuclear and particle physics experiments have increased in complexity in the past decades. This trend occurs especially in high energy physics (HEP) experiments, and concerns not only the increased size and sophistication of detectors and instrumentation, but also the size of experimental collaborations. Particle physics experiments in the 1960's and 1970's involved a relatively small number of researchers; the largest experimental collaborations at the LHC today encompass more than 3000 members.

The average number of members of HEP collaborations at CERN is shown in Fig. 25. It is worthwhile to note that the average values reported in the plots reflect the number of authors in articles published by the experiment, or the current list of members of LHC experiments maintained by CERN; however, the activity of a HEP collaboration usually starts much earlier than the publication of the first articles or the beginning of the data taking period, with the design, construction and commissioning of the detector. For instance, the first LHC experimental teams started their R&D activity in the early 1990's, approximately 20 years before the start of the physics data taking of the experiments. Simulation plays an important role in HEP especially in the early phase, when detectors are designed and optimized. In fact, examination of Fig. 25 shows that the data are consistent with a 10 to 20 year lag between formation of experimental teams and up-ticks in TNS modeling papers. The LCH datum occurs almost exactly 20 years after the step in the TNS data occurring in 1988, and one can observe a small step in the TNS data in connection with the beginning of conceptual design studies for LEP in the late seventies.

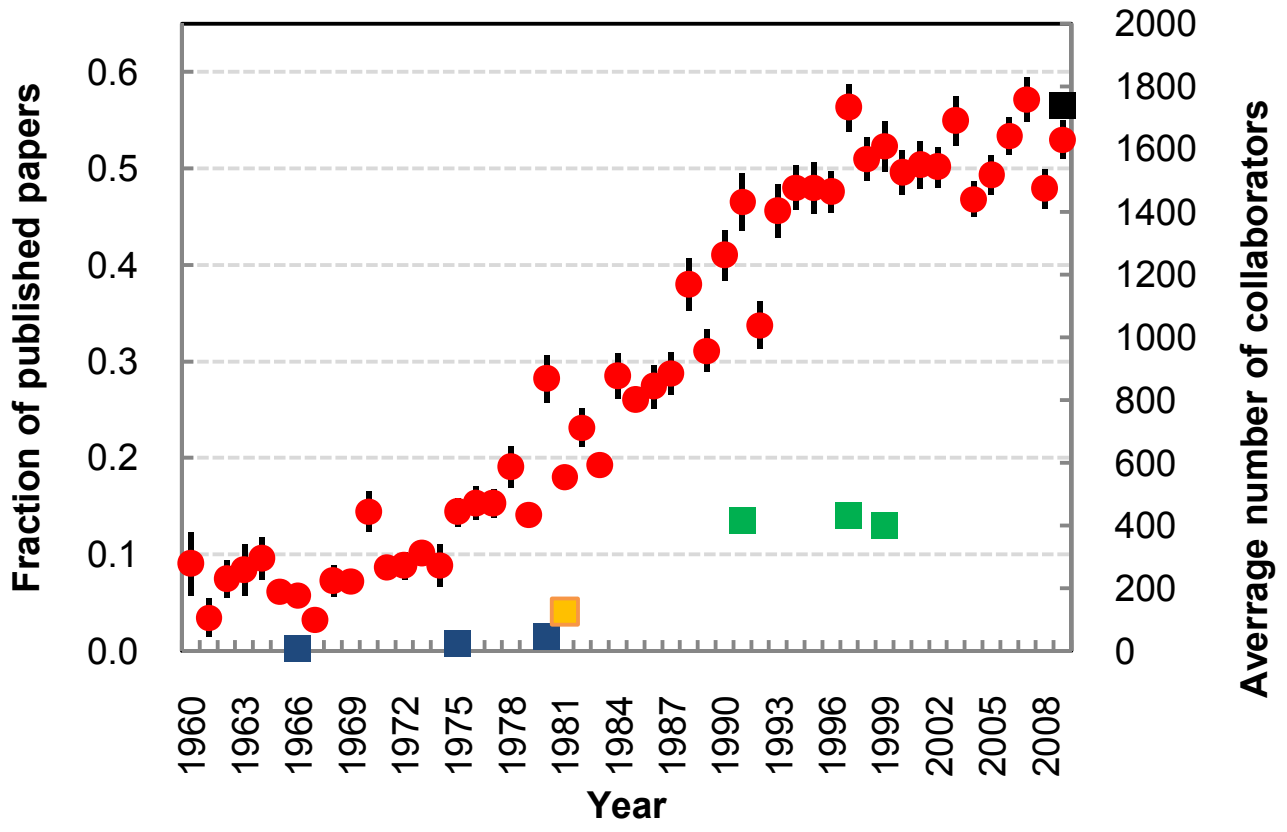

Fig. 25. Average size of experimental collaborations at CERN: fixed target and ISR experiments (blue squares), UA1 (orange square), LEP experiments (green squares), average of LHC experiments (black square), compared to the fraction of TNS papers mentioning "Monte Carlo" or "simulation" (red circles).

Due to this characteristic of the life-cycle of HEP experiments, the data points, which are associated with a single year in Fig. 25, should actually span a longer interval. Keeping this feature in mind, the increase of complexity of HEP experiments, along with the increase of available human resources, could partly explain the trend of increased presence of simulation in the literature.

## V. Conclusion

The presence of simulation has increased in the articles published by several nuclear and particle physics journals over the past fifty years; in instrumentation journals like TNS and NIM, approximately half of the articles mention simulation or Monte Carlo. A scientometric analysis has attempted to identify some technical and socioeconomic factors, which could contribute to the increased popularity of simulation in experimental research.

Gross domestic expenditures in R&D, total employment in science and engineering, number of advanced degrees in a science and engineering fields in various countries, and pro capite income in developed countries show a correlation with the trend of simulation in instrumentation journals. The detailed features of major stock markets are not reflected in simulation publication statistics.

Strong correlation was observed between average income in "first world" countries and the fraction of papers mentioning simulation or Monte Carlo, as well as between gross domestic R&D expenditures and papers mentioning the target terms. It is possible that as the cost of labor increases and the size (or projected size) of collaborations increase, it becomes more economical to invest in simulations than to perform many experiments. However, it cannot be argued that the publication of papers involving simulation or Monte Carlo affects the average income of anyone (except perhaps some of the authors of those papers), because the number of authors of technical papers is miniscule when compared to the number of workers in any country. Nevertheless, it is important to emphasize that correlation does not imply causation.

Other factors have certainly influenced the increased use of simulation in experimental research: the dramatic decrease in the cost of computing, the increased complexity of experiments and detectors, and the availability of general purpose Monte Carlo simulation systems. Although these factors have played an important role in promoting the use of Monte Carlo simulation, their contribution to the observed trend of presence in the literature cannot be directly quantified.